# Effective Focused Crawling Based on Content and Link Structure Analysis

Anshika Pal, Deepak Singh Tomar, S.C. Shrivastava
Department of Computer Science & Engineering
Maulana Azad National Institute of Technology
Bhopal, India
Emails: anshikapal@yahoo.com, deepaktomar@manit.ac.in, scs_manit@yahoo.com

*Abstract* — **A focused crawler traverses the web selecting out relevant pages to a predefined topic and neglecting those out of concern. While surfing the internet it is difficult to deal with irrelevant pages and to predict which links lead to quality pages. In this paper, a technique of effective focused crawling is implemented to improve the quality of web navigation. To check the similarity of web pages w.r.t. topic keywords, a similarity function is used and the priorities of extracted out links are also calculated based on meta data and resultant pages generated from focused crawler. The proposed work also uses a method for traversing the irrelevant pages that met during crawling to improve the coverage of a specific topic.**

*Keywords-focused crawler, metadata, weight table, World-Wide Web, Search Engine, links ranking.*

## I. INTRODUCTION

With the exponential growth of information on the World Wide Web, there is a great demand for developing efficient and effective methods to organize and retrieve the information available. A crawler is the program that retrieves Web pages for a search engine, which is widely used today. Because of limited computing resources and limited time, focused crawler has been developed. Focused crawler carefully decides which URLs to scan and in what order to pursue based on previously downloaded pages information. An early search engine which deployed the focused crawling strategy based on the intuition that relevant pages often contain relevant links. It searches deeper when relevant pages are found, and stops searching at pages not as relevant to the topic. Unfortunately, this traditional method of focused crawling show an important drawback when the pages about a topic are not directly connected. In this paper, an approach for overcoming the limitations of dealing with the irrelevant pages is proposed.

## II. RELATED WORKS

Focused crawling was first introduced by chackrabarti in 1999[1]. The fish-search algorithm for collecting topic-specific pages is initially proposed by P.DeBra et al. [2]. Based on the improvement of fish-search algorithm, M.Hersovici et al. proposed the shark-search algorithm [3]. An association metric was introduced by S.Ganesh et al. in [4]. This metric estimated the semantic content of the URL based on the domain dependent ontology, which in turn strengthens the metric used for prioritizing the URL queue. The Link-Structure-Based method is analyzing the reference-information among the pages to evaluate the page value. This kind of famous algorithms like the Page Rank algorithm [5] and the HITS algorithm [6]. There are some other experiments which measure the similarity of page contents with a specific subject using special metrics and reorder the downloaded URLs for the next crawl [7].

A major problem faced by the above focused crawlers is that it is frequently difficult to learn that some sets of off-topic documents lead reliably to highly relevant documents. This deficiency causes problems in traversing the hierarchical page layouts that commonly occur on the web.

To solve this problem, Rennie and McCallum [19] used reinforcement learning to train a crawler on specified example web sites containing target documents. However, this approach puts burden on the user to specify representative web sites. Paper [20] uses tunneling to overcome some off-topic web page. The main purpose of those algorithms is to gather as many relevant web pages as possible.

## III. PROPOSED ARCHITECTURE

Fig. 1 shows the architecture of focused crawling system [8][9].

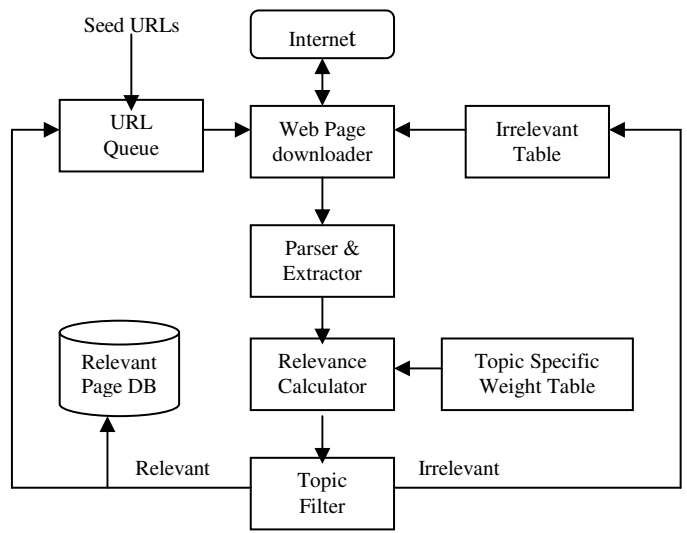

Figure 1.   The architecture of  focused crawler

URL queue contains a list of unvisited URLs maintained by the crawler and is initialized with seed URLs. Web page downloader fetches URLs from URL queue and downloads corresponding pages from the internet. The parser and extractor extracts information such as the terms and the hyperlink URLs from a downloaded page. Relevance calculator calculates relevance of a page w.r.t. topic, and assigns score to URLs extracted from the page. Topic filter analyzes whether the content of parsed pages is related to topic or not. If the page is relevant, the URLs extracted from it will be added to the URL queue, otherwise added to the Irrelevant table.

### A. Topic Specific Weight Table Construction

Weight table defines the crawling target. The topic name is sent as a query to the Google Web search engine and the first K results are retrieved [12]. The retrieved pages are parsed, stop words such as "the" and "is" are eliminated [10], words are stemmed using the porter stemming algorithm [11] and the term frequency *(tf)* and document frequency *(df)* of each word is calculated. The term weight is computed as $w = tf * df$. Order the word by their weight and extract a certain number of words with high weight as the topic keywords. After that weights are normalized as

$$W = \frac{Wi}{Wmax} \qquad (1)$$

Where *Wi* is the weight of keyword *i* and *Wmax* is weight of keyword with highest weight. Table I shows the sample weight table for topic "E-Business".

TABLE I. WEIGHT TABLE

| Keywords | Weight |
|---|---|
| Business | 1.0 |
| Management | 0.58 |
| Solution | 0.45 |
| Corporation | 0.34 |
| Customer | 0.27 |

Now the crawler tries to expand this initial set of keywords, by adding relevant terms that it has intelligently detected during the crawling process [14]. Among the pages downloaded by focused crawler, those which have been assigned a relevance score greater than or equal to 0.9 by equation (3), are so likely to be relevant to the search topic, because the range of relevance score lies between 0 and 1, and the relevancy increases as this value increases, so web pages whose relevance score is over 0.9 is obviously highly relevant pages. Keyword with highest frequency from each of these pages are extracted and added to the table with weight equal to the relevance score of the corresponding page.

### B. Relevance Calculation

#### 1) Page Relevance

The weight of words in page corresponding to the keyword in the table is calculated. The same words in different locations of a page take different information. For example, the title text is more important to express the topic of a page than common text. For this reason, weights are adjusted as follows [8]:

$$f\,kp = \begin{cases} 2 & \textit{title text} \\ 1 & \textit{common text} \end{cases} \qquad (2)$$

*f kp* is the weight of keyword k in different locations of page p. So we can get the overall weight (*wkp*) of keyword k in page p by adding the weights of k in different locations. We have shown an example in fig. 2.

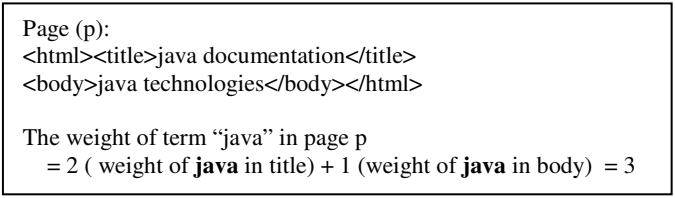

Figure 2. Weighting method

Now we use a cosine similarity measure to calculate the relevance of the page on a particular topic.

$$Relevance\ (t, p) = \frac{\sum_{k \in (t \cap p)} wkt\ wkp}{\sqrt{\sum_{k \in t}(wkt)^2\ \sum_{k \in t}(wkp)^2}} \qquad (3)$$

Here, *t* is the topic specific weight table, p is the web page under investigation, *wkt* and *wkp* is the weight of keyword *k* in the weight table and in the web page respectively. The range of *Relevance (t, p)* lies between 0 and 1, and the relevancy increases as this value increases [15]. If the relevance score of a page is greater than relevancy limit specified by the user, then this page is added to database as a topic specific page.

#### 2) Links Ranking

The Links Ranking, assigns scores to unvisited Links extracted from the downloaded page using the information of pages that have been crawled and the metadata of hyperlink. Metadata is composed of anchor text and HREF information [15].

*LinkScore(j) = URLScore(j) + AnchorScore(j) + LinksFromRelevantPageDB(j) + [ Relevance(p1) + Relevance(p2) +…+ Relevance(pn) ]* (4)

*LinkScore (j)* is the score of link *j*, *URLScore (j)* is the relevance between the HREF information of *j* and the topic keywords, and *AnchorScore (j)* is the relevance between the anchor text of *j* and the topic keywords, we use equation (3) to compute the relevance score. *LinksFromRelevantPageDB(j)* is the number of links from relevant crawled pages to j [8], *Pi* is the ith parent page of URL j [4]. Parent page is a page from which a link was extracted.

### C. Dealing with Irrelevant Pages

Though Focused crawling is quite efficient, it have some drawbacks. From the fig. 3 we can see that at level 2 there are some irrelevant pages (P, Q) which are discarding relevant pages (c, e) at level 3 and (f, g) at level 4 from the crawling path [13].

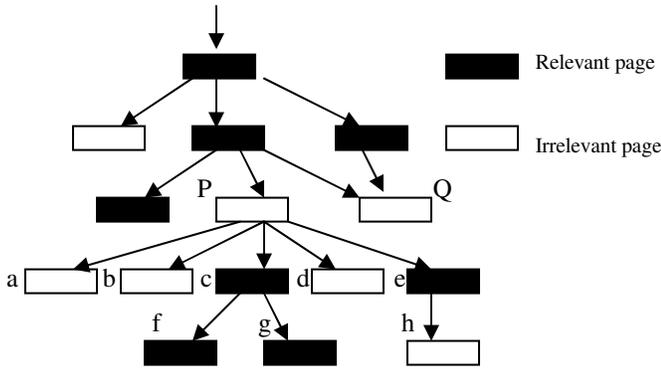

Figure 3.   An example structure of web pages

As a solution of this problem we design an algorithm that allows the crawler to follow several bad pages in order to get a good page. The working principle of this algorithm is to go on crawling upto a given maxLevel from the irrelevant page.

Pseudo code of Algorithm

```
1. if (page is Irrelevant)
2. {
3.     Initialize level = 0;
4.     url_list = extract_urls(page);
               // extract all the urls from the page
5.     for each u in url_list {
6.         compute LinkScore(u) using equation (4);
7.         IrrelevantTable.insert(u, LinkScore(u), level);
               // insert u into irrelevant table with linkscore and level
                  value   }
8.     reorder IrrelevantTable accord to LinkScore(u);
9.     while ( IrrelevantTable.size > 0)
10.    {
11.        get the url with highest score and call it Umax ;
12.        if ( Umax.level <= maxLevel)
13.        {
14.            page = downloadPage(Umax);
                     // download the URL Umax
15.            calculate relevance of the page using equation (3);
16.            if ( page is relevant) {
17.            RelevantPageDB = page;
                     // put page into relevant page database
18.                if ( Umax.level < maxLevel)  {
19.                    level ++ ;
20.                    url_list = extract_urls(page);
21.                    for each u in url_list {
22.                        compute LinkScore(u) using equation (4);
23.                    IrrelevantTable.insert(u, LinkScore(u), level); }
24.                reorder IrrelevantTable accord to LinkScore(u); }
25.            } else {
26.                    for each u in IrrelevantTable  {
27.                        if (LinkScore(u) <= LinkScore(Umax)
                                && u.level == Umax.level)
28.                            u.level ++;   }
29.                }
30.        } }}
```

The main component of the architecture in fig. 1 which help to execute the above algorithm is irrelevant table that allows irrelevant pages to be included in the search path.

In Table II we have shown an example of the above algorithm, for page P in fig. 3.

TABLE II. IRRELEVANT TABLE

| Link | LinkScore | Level |
|------|-----------|-------|
| c | 5 | 0 |
| e | 4 | 0 |
| b | 3 | 0̸  1 |
| a | 2 | 0̸  1 |
| d | 1 | 0̸  1 |
| f | - | 1 |
| g | - | 1 |
| h | - | 1 |

Fig. 3 shows that page P is irrelevant, so accord to the process of line 1-8 in algorithm, URLs that it contains (a, b, c, d, e) are all extracted and inserted into the table with level value 0 and its calculated link score, which are assumed as 1, 2, 3, 4 and 5 for this example, then sort the table (first five entries in table II). Now page c and e are downloaded, and its extracted urls (f, g and h) are added to the table with level value 1 and its corresponding link score (process of line 9-24 and last three entries in table II).

The process of line 25-28 are based on the assumption that if a page is irrelevant and its any child say v is also irrelevant then all the children of this page whose linkscore are less than to v and level value are same as level of v, are less important, so for these urls it would be unnecessary to continue crawling process upto a given maxLevel, we can directly increase level value without downloading these pages, it means that we are reducing the crawling depth of these less meaningful paths.

Now in table II we see that level of page b, a and d are updated from 0 to 1, because when b is downloaded, it seems to be irrelevant and accord to line 27 level of b, a and d are increased.

Clearly, this algorithm can improve the effectiveness of focused crawling by expanding its reach, and its efficiency by reducing the crawling path of less relevant urls so that unnecessary downloading of too many off-topic pages avoided and better coverage may achieved.

IV.   EXPERIMENTAL RESULTS

The experiments are conducted in Java environment. Breadth-First Search (BFS) crawler is also implemented for performance comparison. Target topics were E-Business, Nanotechnology, Politics, and Sports. For each topic crawler started with 10 seed sample URLs, and crawled about one thousand web pages. Google is used to get the seed URLs. The parameter used in this experiment is: Weight Table size = 50, maxLevel = 2, and relevancy limit is equal to the half of the average LinkScore of seed URLs, this value lies between 0.3 to 0.5 in our experiment. Precision metric is used to evaluate the crawler performance.

$$precision\_rate = \frac{relevance\_pages}{pages\_downloaded} \qquad (5)$$

The precision ratios varied among the different topics and seed sets, possibly because of the linkage density of pages under a particular topic or the quality of the seed sets.

Table III shows the final precision rates of four topics after crawling one thousand pages.

TABLE III. THE FINAL RATE OF TOPICS

| Target Topic | Focused Crawler | BFS Crawler |
|---|---|---|
| E-Business | 0.83 | 0.15 |
| Nanotechnology | 0.56 | 0.25 |
| Politics | 0.85 | 0.25 |
| Sports | 0.77 | 0.36 |

We illustrate a crawled result on a two-dimensional graph.

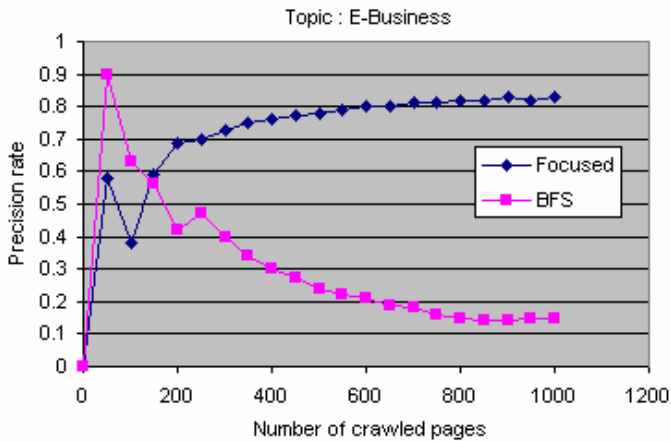

Figure 4. Precision of two crawlers for the topic E-Business

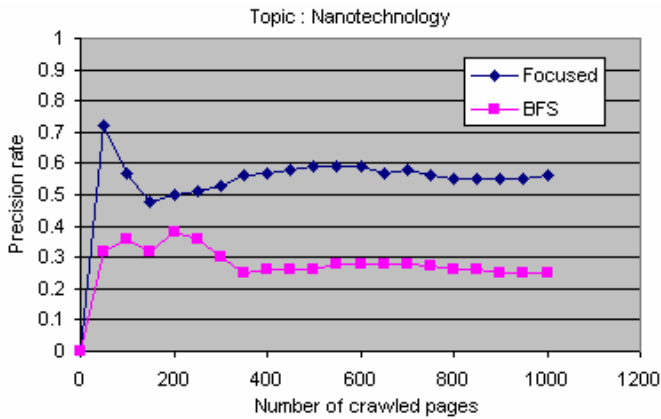

Figure 5. Precision of two crawlers for the topic Nanotechnology

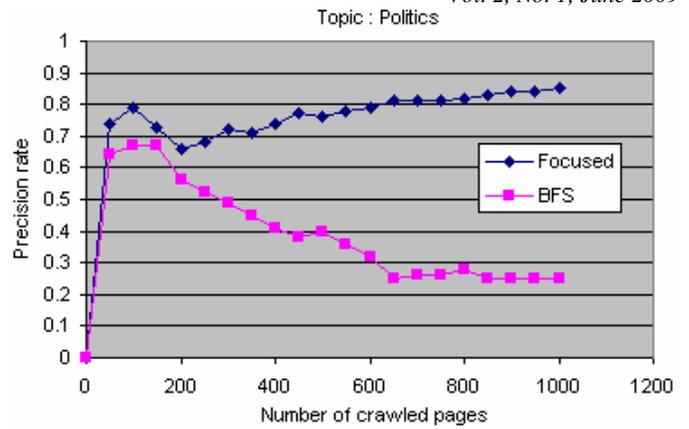

Figure 6. Precision of two crawlers for the topic Politics

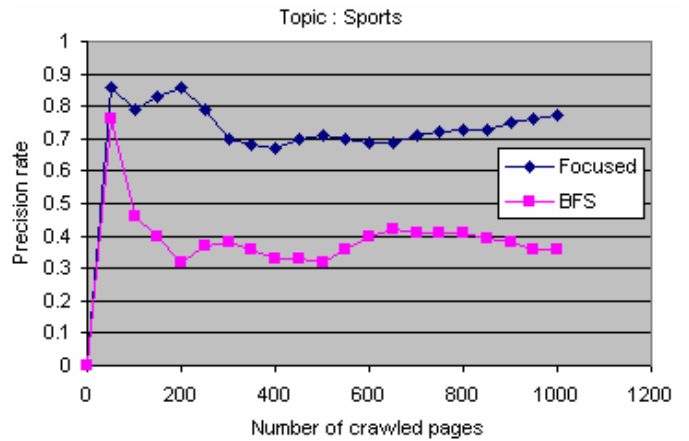

Figure 7. Precision of two crawlers for the topic Sports

Figs. 3, 4, 5, and 6 show the performances of the two crawlers for the topics E-Business, Nanotechnology, Politics, and Sports respectively. These graphs clearly depict that in early stages of the crawling the precision rate of our focused crawler is not so high compared to BFS, but enhancement in the performance appeared after crawling first few hundred pages. It is meaningful since in primary steps of crawling process there is not so much pages in URL queue and so there is some noise. But as the number of downloaded pages increases, the chart would be smoother and precision rate increases.

## V. CONCLUSION & FUTURE WORK

This paper, presented a method for focused crawling that allows the crawler to go through several irrelevant pages to get to the next relevant one when the current page is irrelevant. From the experimental results, we can conclude that our approach has better performance than the BFS crawler.

Although the initial results are encouraging, there is still a lot of work to do for improving the crawling efficiency. A major open issue for future work is to do more extensive test with large volume of web pages. Future work also includes code optimization and url queue optimization, because crawler efficiency is not only depends to retrieve maximum number of relevant pages but also to finish the operation as soon as possible.

## ACKNOWLEDGMENT

The research presented in this paper would not have been possible without our college, at MANIT, Bhopal. We wish to express our gratitude to all the people who helped turn the World-Wide Web into the useful and popular distributed hypertext it is. We also wish to thank the anonymous reviewers for their valuable suggestions.

## AUTHORS PROFILE

**Anshika Pal** Completed B.E. in Computer Science & Engg. and  now pursuing M.Tech degree in Computer Science and Engineering.

**Deepak Singh Tomar** Completed M.Tech & B.E. in Computer Science & Engg. and pursuing PhD in Computer Science & Engg and working as Sr. Faculty in Computer Science & Engg. Department.   Total 14 Years Teaching Experience ( PG & UG ).

**Dr. S. C. Shrivastava** Professor &  Head of the Computer Science & Engg. Department.